\documentclass[10pt,journal,compsoc]{IEEEtran}
\IEEEoverridecommandlockouts
\ifCLASSOPTIONcompsoc   
  \usepackage[nocompress]{cite}
\else
  \usepackage{cite}
\fi

\usepackage[numbers,sort&compress]{natbib}

\usepackage[utf8]{inputenc}
\usepackage{graphicx}
\usepackage{amssymb}
\usepackage{latexsym}
\usepackage{url}
\usepackage{xcolor}
\definecolor{newcolor}{rgb}{.8,.349,.1}

\usepackage{soul}
\usepackage{epsfig}
\usepackage{amssymb}
\usepackage{amsmath}
\usepackage{amsthm}
\usepackage{bm}
\usepackage{amsfonts}
\usepackage{cases}
\usepackage{esint}
\usepackage{graphicx}
\usepackage{subfig}
\usepackage{algorithm} %format of the algorithm
\usepackage{algorithmic} %format of the algorithm
\usepackage{multirow} %multirow for format of Table
\usepackage{amsmath}
\usepackage{xcolor}

\usepackage{makecell}
\usepackage[normalem]{ulem}
\usepackage{overpic}

\newcommand{\gc}[1]{}

\usepackage{array}
\newcommand{\PreserveBackslash}[1]{\let\temp=\\#1\let\\=\temp}
\newcolumntype{C}[1]{>{\PreserveBackslash\centering}p{#1}}
\newcolumntype{R}[1]{>{\PreserveBackslash\raggedleft}p{#1}}
\newcolumntype{L}[1]{>{\PreserveBackslash\raggedright}p{#1}}

\makeatletter
\def\footnoterule{\kern-3\p@
  \hrule \@width 2in \kern 2.6\p@} % the \hrule is .4pt high
\makeatother

\usepackage{enumitem}

\newcommand{\new}[1]{{#1}}

\newcommand{\subsubsec}[1]{\vspace{4pt}\noindent\textbf{#1}}

\providecommand{\keywords}[1]
{
  \footnotesize	
  \textbf{{Keywords---}} #1
}

\begin{document}

\title{\huge
Enabling Design Methodologies and Future Trends for Edge AI: Specialization and Co-design}
\author{Cong Hao$^{1, 2}$, Jordan Dotzel$^3$, Jinjun Xiong$^4$, \\
Luca Benini$^5$, \textit{Fellow, IEEE}, Zhiru Zhang$^3$, and Deming Chen$^1$, \textit{Fellow, IEEE} \\
\small
$^1$Electrical and Computer Engineering, University of Illinois at Urbana-Champaign, \\
$^2$Electrical and Computer Engineering, Georgia Institute of Technology, \\
$^3$Electrical and Computer Engineering, Cornell University \\
$^4$IBM T. J. Watson Research Center, 
$^5$Integrated Systems Laboratory, ETH Zurich \\
callie.hao@gatech.edu, dotzel@cornell.edu, jinjun@us.ibm.com, lbenini@iis.ee.ethz.ch, zhiruz@cornell.edu, dchen@illinois.edu
}

% The paper headers 
\markboth{IEEE Design \& Test}{}

\IEEEtitleabstractindextext{
\begin{abstract}

Artificial intelligence (AI) technologies have dramatically advanced in recent years, resulting in revolutionary changes in people's lives.
Empowered by edge computing, AI workloads are migrating from centralized cloud architectures to distributed edge systems, introducing a new paradigm called edge AI.
While edge AI has the promise of bringing significant increases in autonomy and intelligence into everyday lives through common edge devices,
it also raises new challenges, especially for the development of its algorithms and the deployment of its services, which call for novel design methodologies catered to these unique challenges.
In this paper, we provide a comprehensive survey of the latest enabling design methodologies that span the entire edge AI development stack. We suggest that the key methodologies for effective edge AI development are single-layer specialization and cross-layer co-design.
We discuss representative methodologies in each category in detail, 
including on-device training methods, specialized software design, dedicated hardware design,
benchmarking and design automation, software/hardware co-design, software/compiler co-design, and compiler/hardware co-design. Moreover, we attempt to reveal hidden cross-layer design opportunities that can further boost the solution quality of future edge AI and provide insights into future directions and emerging areas that require increased research focus. 
\end{abstract}

\vspace{12pt}
}

\maketitle
% \IEEEdisplaynontitleabstractindextext
% \IEEEpeerreviewmaketitle

\section{introduction}

\begin{figure*}[t]
    \centering
    \includegraphics[width=1.0\textwidth]{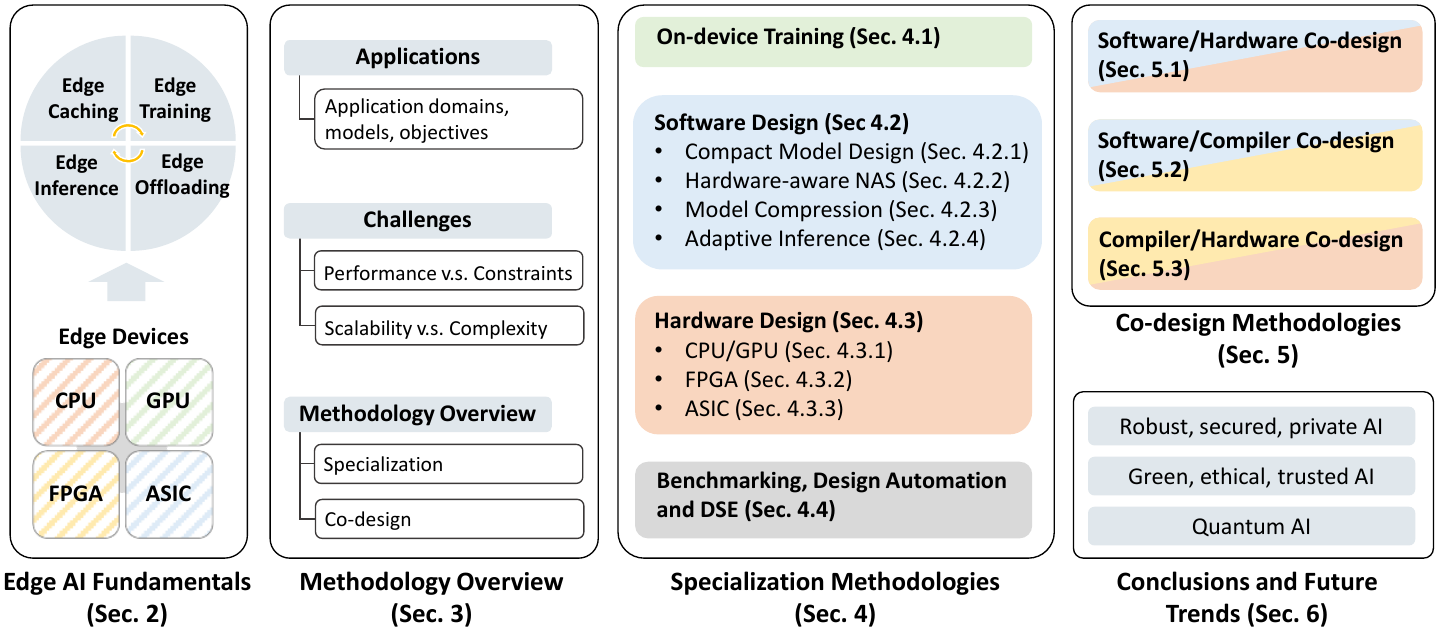}
    \caption{The overall organization of this paper. Section~\ref{sec:fundamentals} introduces edge AI fundamentals. Section~\ref{sec:benefits_challenges} introduces edge AI applications and challenges, and provides an overview of the enabling design methodologies. Section~\ref{sec:sepcialization_methodology} discusses specialization methodologies, while Section~\ref{sec:codesign_methodology} expands on co-design methodologies. Section~\ref{sec:future} concludes the paper and discusses future trends.}
    \label{fig:paper_overview}
\end{figure*}

%% DL important, shift to edge
We are witnessing an unprecedented boom in artificial intelligence (AI), in particular deep learning (DL),
which has made remarkable progress in various areas such as computer vision, natural language processing, health care, autonomous driving, and surveillance.
To accomplish this, AI technologies have broadened from a centralized fashion to mobile or distributed fashion,
opening a new era called edge AI, with dramatic advancements that are substantially changing everyday technology, social behavior, and lifestyles.

%% what is edge intelligence, sets of applications
Edge AI couples intelligence and analysis to a broad collection of connected devices and systems for data collection,
caching, and processing~\cite{shi2016edge}.
It enables a wide variety of new promising applications where data collection and analysis are combined together.
For example, billions of mobile users are exploiting various smartphone applications such as translation services, digital assistants, and health monitoring services.
Meanwhile, the ever-increasing demands of edge AI create tremendous technical challenges, especially for application development and deployment.
Edge devices usually have strict device constraints, such as limited computing capability, memory capacity, and power budget,
often with order-of-magnitude differences from those in server environments.
In addition, as the adoption of edge AI technologies increases, the scale of their applications expands.
The design and deployment for heterogeneous, large-scale edge AI systems and applications require tremendous engineering efforts and introduce complex research challenges.

%% this work focuses on new trends and methodologies
Novel design methodologies and design automation tools are necessary to face the increasing sets of edge AI applications and challenges.
\new{A number of recent works tackle the challenges by spanning the entire development stack, from high-level software/algorithm design to low-level hardware design.
Their enabling design methodologies can be grouped into the two categories, \textbf{specialization and co-design}.
Design specialization carefully adapts common technologies for edge AI application scenarios, such as aggressive model compression or dedicated hardware accelerators. Co-design takes one step further to combine two or more specialized technologies to reveal more optimization opportunities.
}

In this paper, we aim to provide a comprehensive survey of the design methodologies for the new era of edge AI.
There are recent efforts \cite{zhou2019edge, wang2020convergence, xu2020edge} that survey the fundamental concepts and recent advancements of edge AI, but they approach from different angles, such as major domains, applications, and general technologies.
\new{Complementary to these surveys, we discuss Edge AI from a point of view of design methodology. To the best of our knowledge, we are the first to systematically summarize the representative works from different layers (software, compiler, hardware) of the entire develop-deploy stack, and provide a quantitative analysis across layers.
By summarizing related works from different layers and scopes, we hope to motivate cross-layer co-design opportunities that can further contribute to high-quality and flexible edge AI solutions.
Fig.~\ref{fig:paper_overview} provides an overview organization of this work.}

\section{edge ai fundamentals} \label{sec:fundamentals}

In this section, we briefly introduce the four major domains of edge AI, followed by the prevailing edge AI devices, development tools, and frameworks.

\subsection{Major Domains} \label{sec:four_domains}
Edge AI contains four primary domains: edge caching, edge training, edge inference, and edge offloading~\cite{zhou2019edge, wang2020convergence, xu2020edge}.
%\subsubsec{Edge Caching.}
First, \textbf{edge caching} refers to data collection, generation, and storage from edge devices and surrounding environments to support edge applications.
For example, mobile users’ information generated by
themselves is stored in their smartphones,
while environmental monitoring devices and sensors store data at nearby edge servers.
Second, \textbf{edge training} exploits the local data and computing power at the edge, without the high bandwidth requirements of transferring the data first to the cloud. Edge training often requires reduced bit-widths to stay within the device constraints. As an additional benefit, edge training preserves user data privacy by avoiding transferring data first offsite.

Third, \textbf{edge inference} is the execution of AI algorithms at the edge. Considering the network, memory, and computational needs of training, edge inference is often the most common form of edge AI and has received the majority of the recent research attention. 
Finally, \textbf{edge offloading} is a distributed computing scheme
where devices offload their application tasks to the cloud.
Offloading is a promising approach to increase the computation capability of edge devices with limited resources and power budget.

\subsection{Edge Devices}\label{sec:edge_devices}

\begin{figure*}
    \centering
    \includegraphics[width=0.97\textwidth]{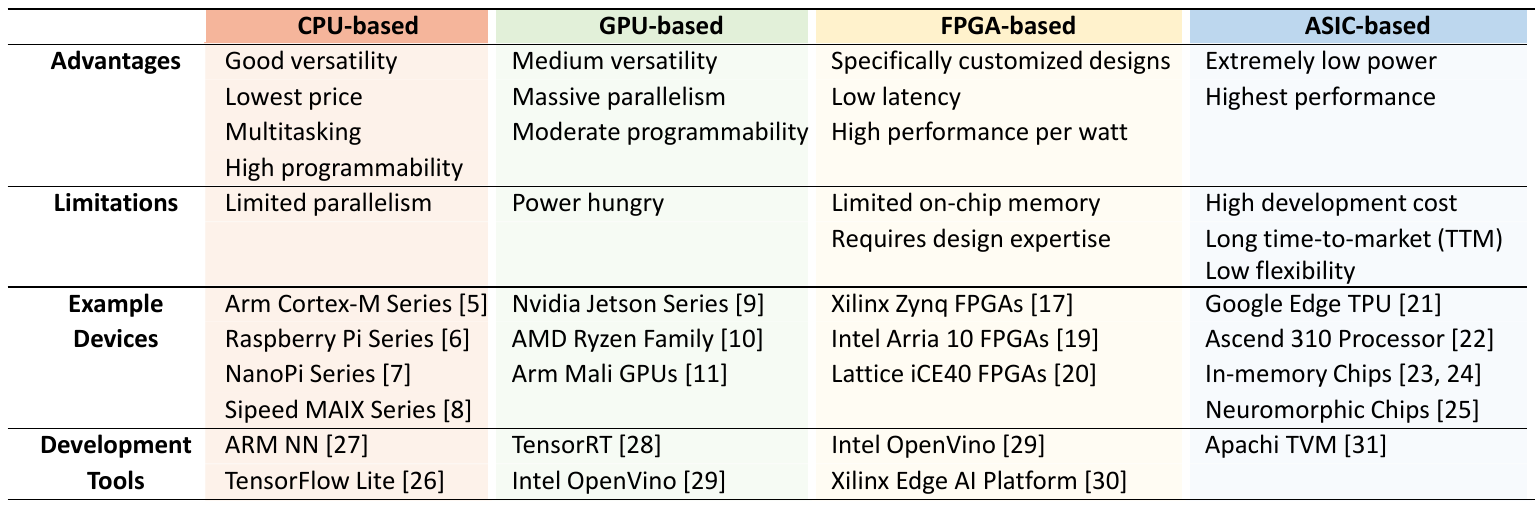}
    \caption{A summary of edge devices with advantages, limitations, example devices, and development tools.}
    \label{fig:edge_devices}
\end{figure*}

The devices that span these major domains are a heterogeneous mix of devices including sensors, monitoring devices, and edge processors, as well as different levels of edge servers, gateways, base stations, and cloud servers.
They can be roughly categorized by their computing class, and the dominant types are CPU-based, GPU-based, FPGA-based, and ASIC-based.
Fig~\ref{fig:edge_devices} provides a brief summary of different edge devices, noting their advantages, limitations, and prominent examples.

\subsubsec{CPU-based.}
CPUs have the longest history and are the most common in many processing and control systems. CPU-based edge devices have the advantages of versatility, low power, low cost, multitasking, and ease of programming.
Since they are primarily designed for general instruction sequences, they usually have limited parallelism. Multi-core architectures improve parallelism, yet CPU-based devices are still not ideal for massively parallel computation. The prevailing CPU-base edge devices include Arm Cortex-M series \cite{arm-cortex}, Raspberry Pi series \cite{raspberrypi}, Nano Pi series \cite{nanopi}, Sipeed MAIX series \cite{Sipeed}, among others.

\subsubsec{GPU-based.}
GPUs were originally designed for image processing and computation with a highly parallel structure, yet now are increasingly used to accelerate AI tasks that require massively parallel computing.
Favoring the high performance through parallelism and programmability, embedded GPUs are specifically designed to be low-power and low-cost, and now are prevalent in many edge devices that accelerate deep learning tasks. Popular GPU-based edge devices include Nvidia Jetson family \cite{nvidia-jetson}, AMD Ryzen family \cite{amd-ryzen}, and Arm Mali GPUs \cite{arm-mali-gpu}.

\subsubsec{FPGA-based.}
FPGAs (field-programmable gate array) are integrated circuits designed to be configured through an array of programmable logic blocks.
They have high application flexibility, for example, supporting arbitrary bit-width and custom numerical formats \cite{rouhani2020mfloat}, and achieve higher performance than CPUs and GPUs in many tasks~\cite{biookaghazadeh2018fpgas, han2017ese, zhang2018dnnbuilder}.
FPGAs usually outperform GPUs on low-latency designs for time-sensitive tasks~\cite{nurvitadhi2017can} and consume lower power compared to CPUs and GPUs while delivering the same throughput.
FPGAs, however, suffer from limited on-chip memory and low programmability.

The leading FPGA vendors, such as Xilinx, Intel, and Lattice, provide a large range of FPGA boards and System-on-Chips (SoCs) targeting various workloads, from power-efficient edge applications to compute-intensive cloud applications~\cite{xilinx-zynq7000, xilinx-spartan7, xilinx-spartan7, intel-arria10, Lattice-iCE40}.

\subsubsec{ASIC-based.}
ASICs (application-specific integrated circuits) can significantly outperform other processors, with a much higher computational throughput and lower power consumption.
Often being massively produced, ASICs take advantage of economies of scale to lower their production price, yet much of these savings are offset by high design and verification costs. Some popular ASICs targeted at edge AI applications include the Google Edge TPU \cite{edge-tpu} and Ascend 310 AI processor \cite{ascend310}.
There are also merging ASIC technologies such as in-memory computing chips~\cite{chen201865nm, wang2020new} and neuromorphic chips~\cite{burr2017neuromorphic}.

\subsubsec{Development Tools \& Frameworks.}
Many design automation tools and frameworks have been developed for edge AI development,
such as: the TensorFlow Lite~\cite{tflite} and the~Arm NN SDK \cite{arm-nn} for CPUs, 
Nvidia TensorRT~\cite{tensorrt} for GPUs, Intel OpenVINO~\cite{intel-openvino} for heterogeneous edge devices, the Xilinx Edge AI Platform \cite{xilinx-edge-ai} for embedded-CPU-based FPGAs, and the Apache TVM \cite{tvm} for CPUs, GPUs, and specialized ASICs. 
\begin{figure}
    \centering
    \includegraphics[width=0.47\textwidth]{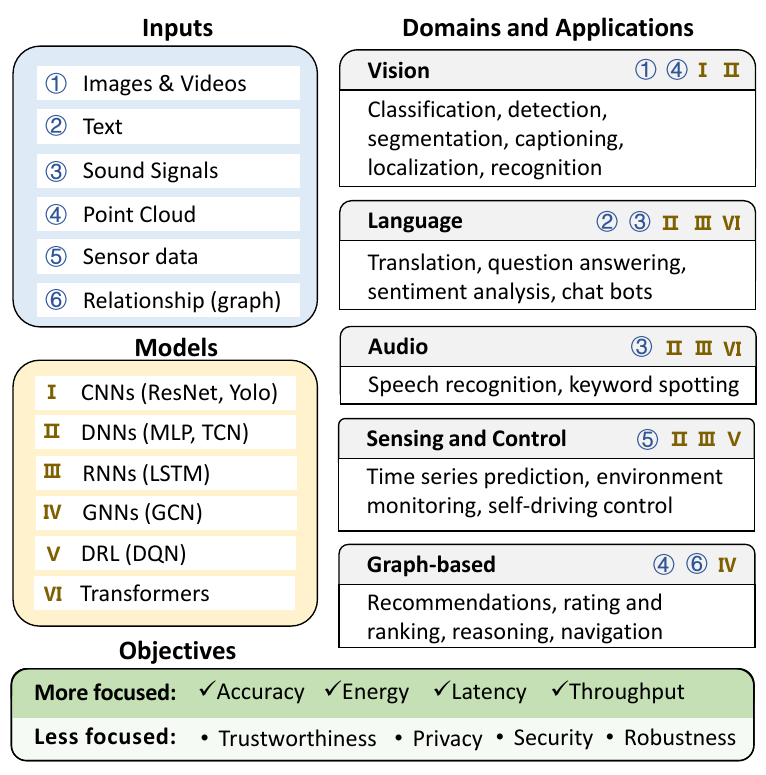}
    \caption{AI applications and models.}
    \label{fig:AI_applications}
\end{figure}

\begin{figure}
    \centering
    \includegraphics[width=0.47\textwidth]{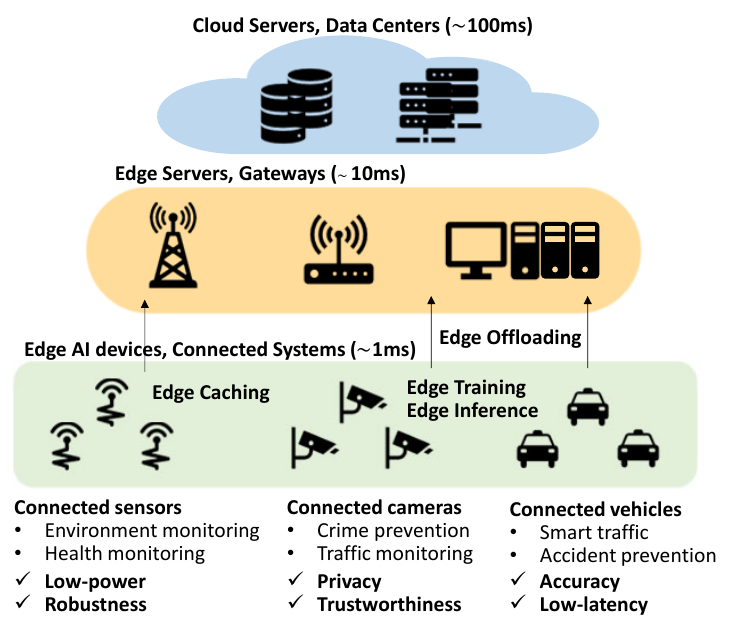}
    \caption{Edge AI use-case diagram.}
    \label{fig:use_case_diagram}
\end{figure}

\section{enabling methodology overview} \label{sec:benefits_challenges}

The widespread adoption of edge AI across different domains is currently limited by its challenging development process, which requires novel design methodologies.
\new{In this section, we discuss the general problem setup of edge AI by introducing
prevailing applications and models, development challenges of edge AI technologies, and then provide an overview of enabling methodologies.}

\subsection{Applications and Challenges} \label{sec:applications}

\subsubsec{Applications.}
\new{
Fig.~\ref{fig:AI_applications} provides a brief summary of prevailing edge AI applications with their most common inputs, models, and optimization objectives.
% Common application domains include vision, language, audio, sensing and control, and graph-based applications, while commonly used AI models include convolutional neural networks (CNNs), deep neural networks (DNNs), recurrent neural networks (RNNs), graph neural networks (GNNs), deep reinforcement learning (DRL), etc.
% At present, many research interests in edge AI fall in the category of vision and language, taking the inputs of images and texts for CNNs, DNNs, and RNNs and more focusing on model accuracy, energy, latency, and memory footprint.
In particular, one commonly accelerated class of models, GNNs, have gained more attention in recent years. There is also a growing need for trustworthy, private, and secured edge application developments.}

\new{The diagram in Fig.~\ref{fig:use_case_diagram} examples a typical AI system, spanning from the cloud server, edge server, to the end-to-end edge devices, with a typical latency requirement of 1ms, 10ms, 100ms, respectively~\cite{xiong2020challenges}.
Three use cases at the edge are presented.
The left set of edge devices are environmental and health sensors that collect at the edge and process data at nearby edge servers to reduce the memory footprint and computation. This is an example of edge offloading, and the most important design metrics are often low-power and robustness.
The middle set of devices are cameras for crime prevention and traffic monitoring, which may require edge training and edge inference to adaptively adjust the current detection targets. Edge offloading may also be necessary to obtain higher quality results. 
The key design metrics with connected cameras are often privacy and trustworthiness.
Finally, the rightmost devices are autonomous vehicles that involve edge training, inference, and offloading. In this case, accuracy and computation latency are more important for making safe decisions on the road.
}

\subsubsec{Challenges.} \label{sec:challenges}
To successfully design and deploy solutions for those applications, many challenges must be overcome.
The largest challenge is achieving high performance under strict device constraints, which include limited computing capability, memory capacity, and power budget.
For instance, the battery life for a personal drone is only 20 to 30 minutes, where only a small portion of power (less than $5\%$) can be allocated for computing and processing system~\cite{boroujerdian2018compute}. 
Another challenge is the complexity of the application at scale.
Large-scale AI applications contain numerous edge nodes, e.g., a smart city may contain millions of surveillance cameras and sensor nodes. In addition, with complexity comes heterogeneity. An IoT application may contain largely different devices with varying resources, computing capability, memory capacity, and power budgets.
% Meanwhile, there are largely diverse applications, which differ in domains, scales, and targeted devices.

\subsection{Design Methodology Overview}

\new{
Driven by the broad applications and challenges, edge AI technologies are largely empowered by rapidly improving design methodologies. 
The development and deployment stack for edge AI contains multiple layers: high-level software development, model training, compiler, and low-level hardware deployment.
Their design methodologies can be summarized into two categories, single-layer \textbf{specialization} and cross-layer \textbf{co-design}:}

\subsubsec{Specialization Methodologies:}
\begin{enumerate}
    \item {
\new{Training: low-precision and on-device training methods, to develop lightweight AI models and enable on-device training.}
    }
    \item {
\new{Software: edge-oriented inference algorithm design methods, to assure high-quality algorithms.}
    }
    \item {
\new{Hardware: accelerator design methods, to achieve low latency and low power.}
    }
    \item {
\new{Benchmarking, automation, and design space exploration (DSE): to effectively evaluate and improve design quality as well as productivity.}
    }
\end{enumerate}

\subsubsec{Co-design Methodologies:}
\begin{enumerate}
    \item{
\new{Software/hardware co-design: mutually specialized software model and hardware accelerator, to simultaneously deliver accurate models and accelerators.}
    }
    \item {
\new{Software/compiler co-design: software model design with compiler customization, to enable efficient on-device execution.}
    }
    \item {
\new{Compiler/hardware co-design: specialized hardware architecture and the supporting compiler and programming library, to deliver efficient accelerators with efficient executions.}
    }
\end{enumerate}

\begin{figure*}
    \centering
    \includegraphics[height=0.9\textheight]{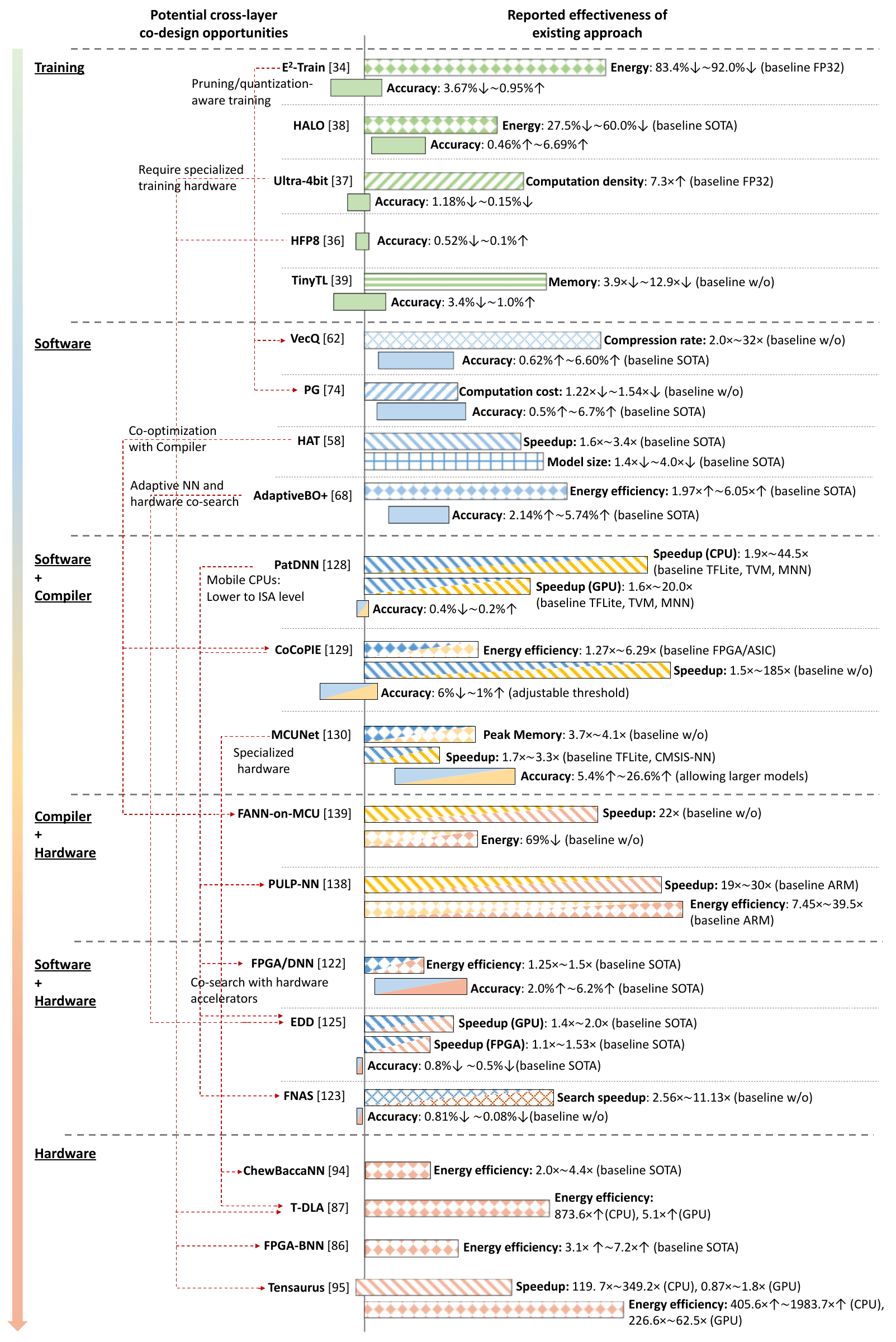}
    \caption{An holistic analysis of representative works from different stages in the development-deploy stack. The performance data are shown in ranges as reported in the proposed solutions. For example,
    Accuracy: $3.67\%\downarrow \sim 0.95\%\uparrow$ represents the range of accuracy changes due to the proposed solution.
    Due to the different scales and design metrics, the lengths of the bars are normalized categorically (energy efficiency, accuracy, etc.) based on their mean values within the range. }
    \label{fig:all_work_quantitative}
\end{figure*}

In the following sections, we provide a comprehensive review of related works based on their dominant design methodology: specialization (Sec.~\ref{sec:sepcialization_methodology}) or co-design (Sec.~\ref{sec:codesign_methodology}).
\new{Moreover, in Fig.~\ref{fig:all_work_quantitative}, we analyze the effectiveness of these methodologies and demonstrate their  benefits, discuss our observations, and provide insight where possible.
By summarizing related works from various layers and scopes all together, we indicate potential optimization opportunities that can further contribute to high-quality and flexible edge AI solutions.}

\section{specialization methodologies}
\label{sec:sepcialization_methodology}

\begin{figure}
    \centering
    \includegraphics[height=0.95\textheight]{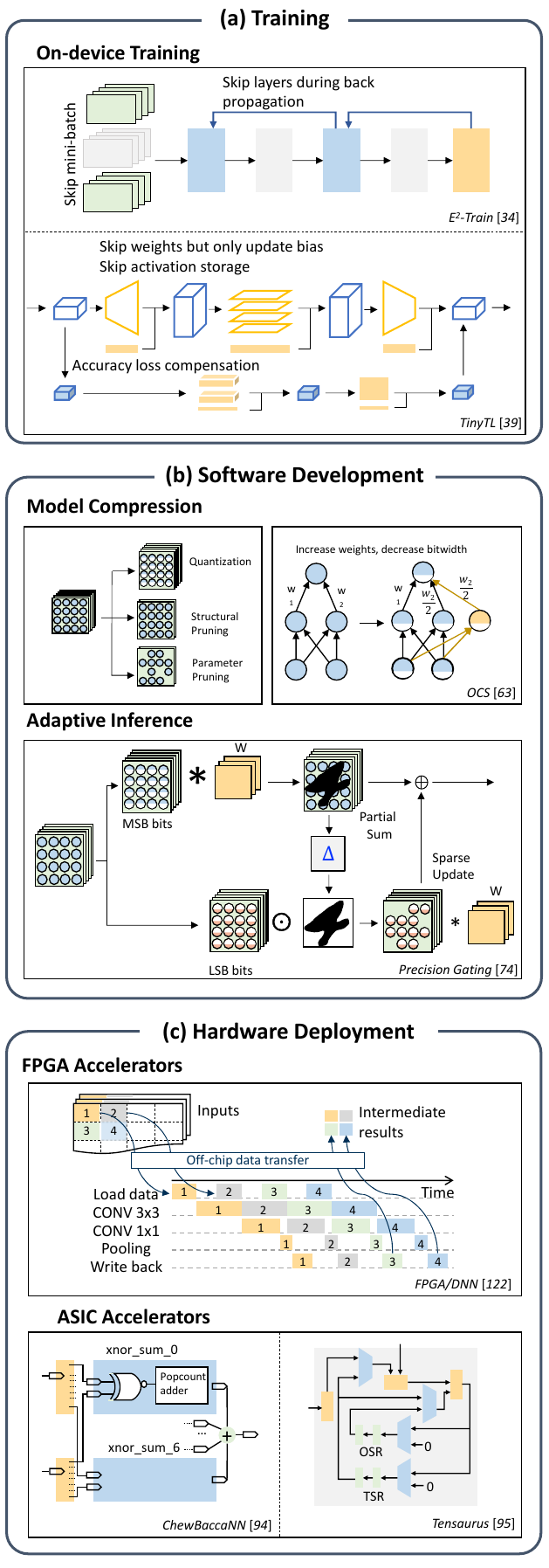}
    \caption{Examples of specialization methodologies: on-device training, software development, and hardware deployment.}
    \label{fig:sp_design_all}
\end{figure}

\new{
Highly specialized designs and technologies are the foundations of successful edge AI development and deployment.
Fig.~\ref{fig:sp_design_all} provides several representative examples for edge training, software development, and hardware deployment.}
\new{
For on-device training, one important idea is to intelligently ``skip'' uncritical operations to pursue energy, latency, and memory reduction (Fig.~\ref{fig:sp_design_all} (a)).}
\new{
For software development, in addition to model compression and pruning, adaptive inference is a promising technique where different models are dynamically chosen based on varied task requirements (Fig.~\ref{fig:sp_design_all} (b)).}
\new{
For hardware deployment, especially FPGAs and ASICs, there are multiple levels of specialization, from the architectural level to the gate level (Fig.~\ref{fig:sp_design_all} (c)). 
}

\subsection{Low-precision and On-device Training}

\new{
Low-precision and on-device edge training can be optimized by varying different components, such as training algorithm, data type, optimizer.}
\new{
For instance, E$^2$-Train~\cite{wang2019e2} and FracTrain~\cite{fu2020fractrain} aim to reduce on-device training energy from the algorithmic level. E$^2$-Train proposes stochastic mini-batch dropping, selective layer update, and sign prediction;
FracTrain proposes to gradually increase the precision of activations, weights, and gradients.}
\new{Other works propose different data type formats during training.
HFP8~\cite{sun2019hybrid} proposes a hybrid FP8 format that uses different floating point formats during forward and backward propagation, achieving negligible accuracy degradation.
Ultra-4bit~\cite{sun2020ultra} further reduces the bit-width down to 4-bit by proposing adaptive gradient scaling to deal with the insufficient data range and resolution in quantized gradients.}
\new{
HALO~\cite{li2020halo} proposes a practical meta-optimizer dedicated to resource-efficient on-device adaptation by introducing a new regularizer that reduces data size or training iterations to reach a specified accuracy.}
\new{
To reduce the memory footprint on edge devices during fine-tuning, TinyTL~\cite{cai2020tinytl} proposes freezing the weights and only learns the bias modules.}

Recently, Federated Learning (FL)~\cite{mcmahan2017communication, bonawitz2019towards} has emerged as a promising distributed training paradigm at the edge. Yang et al.~\cite{yang2019federated} provide a comprehensive survey on FL, so we omit the detailed discussions in this paper.

\subsubsec{Observations and Insights.}
\new{
Various works of on-device training can be potentially combined to achieve better performance. For example, TinyTL may be seamlessly integrated into other training algorithms for fine-tuning, while FracTrain and Ultra-4bit may be complementary to each other as they both exploit progressive scaling for gradients.
}

\new{
Notably, most novel training techniques require specific hardware support. For example, both HFP8 and Ultra-4bit require specially designed floating point units (FPUs);
FracTrain assumes BitFusion~\cite{sharma2018bit} architectures to support dynamic precision, but there is still no dedicated training hardware.
While Wang et al. \cite{wang2020new} propose an MRAM-based process in-memory accelerator for floating point DNN training, more training hardware design is expected.
}

\subsection{Software Design} \label{sec:software_develop}

Powerful AI applications require increasingly more computing resources, which complicates their deployment on edge devices. 
Specialized software models, including compact and hardware-aware model design, compression, and adaptive inference, are essential to building edge systems that perform under strict constraints.

\subsubsection{Compact Model Design}

\new{
Many works focus on directly designing lightweight AI models, each of which applies different methods for sparsifying the model. These include the SqueezeNet~\cite{iandola2016squeezenet}, MobileNet family~\cite{howard2017mobilenets, sandler2018mobilenetv2, howard2019searching}, the ShuffleNet family~\cite{zhang2018shufflenet, ma2018shufflenet}, and ChannelNets~\cite{gao2020channelnets}.
SqueezeNet aggressively reduces the number of channels through $1\times 1$ convolutions, while MobileNet decomposes the standard convolution into a depth-wise separable convolution followed by a $1\times1$ point-wise convolution.
ShuffleNet reduces the convolution complexity through channel grouping,
while ChannelNets further proposes channel-wise, group channel-wise, and depth-wise separable channel-wise convolutions.}

More recently, other works also suggest model structures specifically optimized for efficient inference. EEG-TCNet~\cite{ingolfsson2020eeg} proposes a temporal convolutional network (TCN) that achieves impressive accuracy while requiring few trainable parameters. HadaNet~\cite{zhao2019ugconv} proposes a Hadamard variant of the ShuffleNet~\cite{zhang2018shufflenet}, which more efficiently mixes channel information. Further, BFT~\cite{vahid2020bft} extends this line of research and uses the butterfly transform to perform channel fusion, finding increased performance especially in very small networks.

\subsubsection{Hardware-aware Model Design}\label{sec:hardware_aware_NAS}
In contrast to handcrafted models by machine learning experts, a more recent approach is neural architecture search (NAS) to automatically design the model architecture to achieve outstanding performance.
\new{
Extended from compact model designs, the work SqueezeNext~\cite{gholami2018squeezenext} proposes a new family of SqueezeNet-like architectures, whose designs were guided by simulation results on a neural network accelerator.
MNasNet~\cite{tan2019mnasnet} targets on-device latency within its NAS objective, leading to execution speedup on edge devices such as mobile phones.
Similarly, MobileNetV3~\cite{howard2019searching} proposes hardware-aware NAS accompanied by the NetAdapt~\cite{yang2018netadapt} algorithm.}

Most recently, SkyNet~\cite{zhangskynet} uses a different bi-directional DNN design approach with a comprehensive understanding of the hardware constraints, demonstrating its effectiveness by winning the System Design Contest for low power object detection in the 56th IEEE/ACM Design Automation Conference (DAC-SDC).
Besides computer vision, hardware-aware NAS also demonstrates success in natural language processing (NLP) tasks.
For example, a recent work HAT~\cite{hanruiwang2020hat} proposes hardware-aware transformers that exploit NAS to discover efficient models on various devices.

\subsubsec{Observations and Insights}
\new{
The great success of NAS and hardware-aware NAS suggests that it could be a promising tool to assist future AI edge designs. For example, NAS has been applied to defend against adversarial attacks~\cite{chen2020anti} and ensure secure inference~\cite{bian2020nass}.
While these works are in their early stage, they reveal a potential solution to edge AI robustness, security, privacy, and trustworthiness.
}

\new{One disadvantage is that NAS uses a lot of computational resources and power. 
On the one hand, NAS can be a powerful tool to search for light-weight and compact AI models; on the other hand, the search process itself may have a large carbon footprint~\cite{strubell2019energy}. Designing environment-friendly NAS algorithms is as important as designing hardware- and environment-friendly models.
}

\subsubsection{Model Compression} \label{sec:model_compression}
Model compression includes a series of techniques to reduce the model size, allowing for lower latency, power, and potential area reduction.
Within model compression, quantization has become standard practice in deploying models at the edge.
For instance, VecQ~\cite{gong2020vecq} introduces a DNN quantization solution, which utilizes a vector loss rather than traditional L2 loss as the quantization objective function to achieve higher model accuracy. 
Likewise, outlier channel splitting (OCS)~\cite{zhao2019ocs} devises an extension to existing quantization methods to handle weight outliers to reduce quantization accuracy loss without DNN retraining. 

In addition to general model compression methods, several techniques target specific edge devices such as low-power microcontroller units (MCUs).
For example, Rusci et al.~\cite{rusci2019memorydriven} propose a mixed-precision quantization on microcontrollers, while Capotondi et al.~\cite{capotondi2020cmix} propose an open-source mixed low-precision inference library called CMix-NN for quantized networks on edge MCUs.

\subsubsection{Adaptive Inference} \label{sec:adaptive_inference}

\new{
Adaptive inference is another promising technology that can effectively reduce the model execution latency and energy consumption. The core idea is to adapt models during inference based on the complexity of the current task.
So far, the model adaption happens at least along four common dimensions: number of layers or cascaded models~\cite{panda2016conditional, jayakodi2018trading, stamoulis2018designing}, number of channels~\cite{yu2018slimmable, chin2020pareco, jayakodi2020design, hua2019channel}, input image resolution~\cite{chin2019adascale}, and computation precision~\cite{zhang2020pg, zhang2020fracbnn}.
}

\new{
Along the layer dimension, Panda et al.~\cite{panda2016conditional} propose a conditional deep learning (CDL) network, which can identify the variability in the difficulty of input instances and conditionally activate the deeper layers of the network.
Extended from CDL~\cite{panda2016conditional}, Jayakodi et al.~\cite{jayakodi2018trading} propose on-the-fly classifier selection: simple classifiers for easy inputs and complex classifiers for hard inputs. 
Stamoulis et al.~\cite{stamoulis2018designing} propose a systematic approach for hyper-parameter optimization of adaptive CNNs, using Bayesian optimization to determine the number of channels, kernel sizes, and the number of units in the fully connected layers.}

\new{Along the channel dimension, SlimNets~\cite{yu2018slimmable} is an early work that allows the DNN to adjust the number of channels according to on-device benchmarks and resource constraints with switchable batch normalization.
LEANets~\cite{jayakodi2020design} further allows finer energy-accuracy tradeoffs by introducing confidence threshold vectors.
The work PareCO~\cite{chin2020pareco} proposes varied width multipliers for different layers and pursues Pareto solutions between theoretical speedup and accuracy.
The channel gating (CG)~\cite{hua2019channel} uses a base set of channels and a conditional set, and determines whether to compute the additional channels based on learnable thresholds.
-}

\new{Instead of changing the model structure, some works focus on the resolution and precision of the model. Adascale~\cite{chin2019adascale} proposes to dynamically change the input image resolution to improve both accuracy and speed for video object detection tasks.
Precision gating (PG)~\cite{zhang2020pg} uses a dual-precision mode for edge inference that dynamically selects which features to run in half and full precision.}

\subsubsec{Observations and Insights.}
\new{Adaptive inference is a relatively independent approach, thus could be integrated with other techniques to pursue larger benefits. First, different dimensions of adaptive inference can be jointly considered, such as channel dimension, precision dimension, and layer dimension. 
Second, NAS techniques may help discover better adaption strategies and network architectures. 
Third, similar to on-device training, the pre-determined or learned adaption strategy may drift when input data change; therefore, adaptive inference may also require on-line training and adjustment.
Adaptive inference can be expected to be combined with software/hardware co-design to improve performance. Hua et al.~\cite{hua2019boosting} advances this line of work by proposing an RTL accelerator for channel gating.
}

\subsection{Hardware Deployment} \label{sec:hardware_deploy}

The hardware must also be specialized for edge AI applications. Currently, these applications often use devices from one of the following categories: CPU, GPU, FPGA, and ASIC.
Among them, CPU and GPU do not allow users to alter the hardware design,
%but only optimizations on the programming library level, 
while FPGA and ASIC provide more flexibility so designers can customize the designs to meet particular requirements.
FPGA and ASIC designs can both optimize at architectural level, processing element level, and register level, while ASIC designs can further customize at the gate level.

\subsubsection{CPU-/GPU-based}
Since CPUs and GPUs provide very limited flexibility for designers to alter hardware architecture, most of the deployments focus on software optimizations.
Nevertheless, there are opportunities for efficient memory management and power control. The work Q-EEGNet~\cite{schneider2020q} proposes algorithmic and implementation optimizations for EEGNET~\cite{lawhern2018eegnet}, a compact convolutional neural network (CNN) on an ultra-low power RISC-V based System-on-Chip, to achieve significant speedup and reduction of memory footprint.
Di et al.~\cite{di2020idleness} discuss the development of a power management infrastructure based on the Dynamic Voltage and Frequency Scaling (DVFS), significantly saving power without penalty to the execution performance.

\subsubsection{FPGA-based}

With high flexibility of bit-width manipulation and data flow customization, FPGA accelerators are extremely suitable for quantized, sparse, hybrid, or customized neural networks~\cite{zhang2017machine, zhuge2018face, papakonstantinou2011multilevel} with the high-level synthesis (HLS) and automated compilation tools~\cite{rupnow2011study, papakonstantinou2009fcuda, chen2009lopass}.
For edge applications, Zhao et al.~\cite{zhao2017accelerating} are the first to study FPGA acceleration for binarized neural networks (BNNs), T-DLA~\cite{chen2019t} is the first instruction-based accelerator for ternarized neural networks (TNNs),
and Wang et al.~\cite{wang2018design} support hybrid low bit-width quantized DNNs.
For sparse and hybrid DNNs, Huang et al.~\cite{huang2019accelerating} propose a configurable inference engine capable of processing different sizes of sparse DNNs, while HybridDNN~\cite{ye2020hybrid} has a hybrid architecture composed of spatial/Winograd convolution processing elements.
Additionally, Carreras et al.~\cite{carreras2020optimizing} present an enriched architectural template supporting efficient TCNs, together with an algorithm for optimal execution/scheduling of data-transfers to boost the implementation performance.

\subsubsection{ASIC-based}

Compared with the other solutions, ASICs usually achieve higher performance, smaller area, and orders of magnitude of better energy efficiency.
The DianNao family~\cite{chen2016diannao} is an early series of machine learning acceleration chips in academia.
Eyeriss~\cite{chen2016eyeriss} is a representative work for reconfigurable CNN acceleration, composed of an array of processing elements (PEs) with a reconfigurable on-chip network.
ChewBaccaNN~\cite{andri2020chewbaccann} is a recent binary CNN acceleration chip, which especially optimizes at gate-level.

Not limited to DNN designs, accelerators for common operations, such as tensor-based and matrix-based multiplications, are also of great importance.
One recent work, Tensaurus~\cite{srivastava2020tensaurus}, is a hardware accelerator capable of both sparse and dense tensor computation and other common mixed sparse-dense matrix operations.

\subsubsec{Observations and Insights.}
\new{
Efficient hardware designs are the foundation of high-quality edge AI solutions.
Almost every design approach requires dedicated hardware consideration, such as low-precision training, quantization, structured or non-structured pruning, hardware-aware model design, and adaptive inference.
Ignoring hardware considerations may result in large gaps between theoretical and actual benefits. 
For example, a study by Ma et al.~\cite{ma2019non} points out that non-structured pruning without dedicated hardware is even considered harmful. 
Therefore, hardware design should always be considered when developing the application, and in Section~\ref{sec:codesign_methodology}, we discuss hardware-related co-design methodologies in detail.
}

\subsection{Benchmarking, Design Automation and DSE}

Edge AI development requires intensive domain knowledge and engineering efforts. It comes with many optimization parameters that result in an extremely large design space.
Second, future edge AI systems bring in scalability problems and make it time-consuming to search for high-quality solutions.
Thus, informative benchmarking, design automation, and DSE tools are indispensable for rapid and high-quality edge AI development.

\subsubsection{Scalable Benchmarking of Models, SW, and HW}

Scalable benchmarking offers many combinations of different design metrics, such as accuracy, performance, power/energy, and cost, that enable quick iteration and evaluation of designs. A recent work, titled MLModelScope~\cite{dakkak2018mlmodelscope, li2020design}, proposes a scalable benchmark platform mainly for deep learning based workloads.
MLModelScope supports different combinations of DL models, frameworks, and hardware devices, allows scalable evaluation, and reports informative benchmarking results.
On GPU devices, MLModelScope further provides an automated analysis tool, called XSP~\cite{li2019acrossstack}, to build an integrated view of various performance-related metrics of workloads across the entire stack.
Another recent benchmarking platform is called MLCommons~\cite{MLCommons}, which provides benchmarking, datasets, and practical innovative ML models.

\subsubsec{Observations and Insights.}
\new{
An interesting yet challenging future direction is to extend benchmarking platforms to better support edge devices with higher flexibility, such as ASICs and FPGA devices.
One major challenge is to incorporate hardware design flexibility while still providing the same level of abstraction for software tools, i.e., when users actively change the target hardware architecture, the benchmarking tools must capture the new features and provide accurate performance reports.
Also, the software tools for edge devices are much more diverse than those for servers and clouds, and the commonly accepted software interfaces (such as Tensorflow, PyTorch, and MxNet) and capabilities
(such as CUDA, cuBlas, and cuDNN \cite{cudnn}) are still lacking. There are no well-developed profiling and tracing tools for edge devices either, such as VTune~\cite{vtune} for CPUs, and Nsight~\cite{nsight} and CUPTI~\cite{cupti} for GPUs.
}
 
\subsubsection{Design Automation and Design Space Exploration}

Design automation and DSEs can be applied on different devices and platforms including CPUs, FPGAs, ASICs, and System-on-Chips (SoCs). 
Targeting embedded CPUs, De et al.~\cite{de2020automated} apply reinforcement learning for DSE at different levels of the software stack to search for optimized solutions.
Targeting FPGA platforms,
DNNBuilder~\cite{zhang2018dnnbuilder} is a representative automation and DSE tool including RTL DNN templates, a layer-based pipeline architecture, and an automatic DNN accelerator generator. 
Meanwhile, the tool LAMDA~\cite{ustun2019lamda} speeds up the configuration tuning during FPGA design flow by automatically selecting the configuration options.
Targeting both FPGAs and ASICs, AutoDNNChip~\cite{xu2020autodnnchip} 
first predicts DNN accelerator performance such as area and throughput, applies DSE search for optimal configurations, and then automatically generates synthesizable RTL code. 
Targeting SoC platforms, Zuo et al.~\cite{zuo2015polyhedral} first present an automated SystemC generation and DSE flow,
which converts a subset of C/C++ into a full SystemC description through polyhedral models and generates Pareto-optimal solutions.
A follow-up work, named RIP~\cite{zuo2017accurate}, is proposed to enable hardware acceleration by utilizing hardware/software partitioning to minimize the overall program latency. 

\subsubsec{Observations and Insights.}
\new{
Design automation and DSE enable highly productive edge AI development by breaking the boundaries between different layers of the development stack, opening more opportunities for co-design without requiring domain-specific knowledge. For example, an automated FPGA~\cite{zhang2018dnnbuilder} or ASIC development tool~\cite{xu2020autodnnchip} can be directly used by software designers to evaluate their models.
Vice versa, a well-established low-precision training and quantization framework can be very helpful for hardware developers. Ultimately, design automation and DSE tools can significantly improve design productivity, contribute to the ecosystem of edge AI, and enable open-source software and hardware development.
}
\section{co-design methodologies} 
\label{sec:codesign_methodology}

\begin{figure}
    \centering
    \includegraphics[height=0.93\textheight]{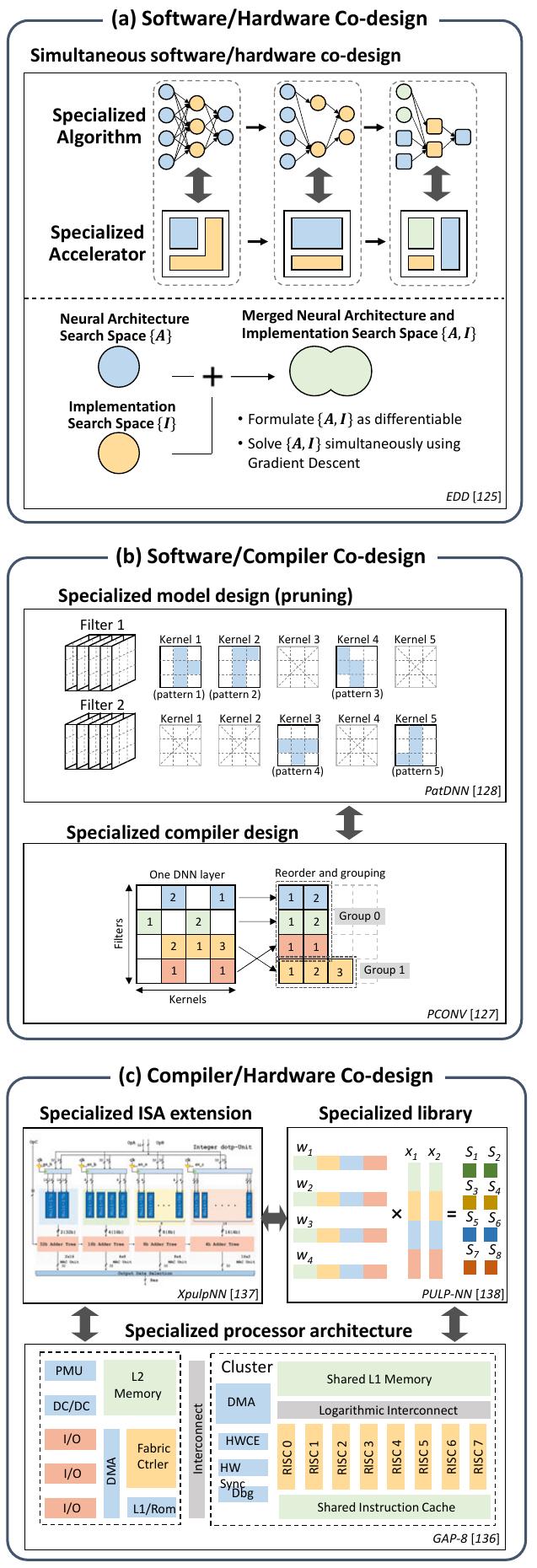}
    \caption{Examples of co-design methodologies: software/hardware co-design, software/compiler co-design, and compiler/hardware co-design.}
    \label{fig:co_design_all}
\end{figure}

\new{
While specialization for individual design layers has led to accurate and effective Edge AI, optimizing multiple layers concurrently, i.e. cross-layer \textbf{co-design} unleashes more opportunities.
In this paper, we summarize three major co-design methodologies, as shown in Fig.~\ref{fig:co_design_all}: software/hardware, software/compiler, and compiler/hardware.
Software/hardware co-design aims at simultaneously designing DL models and accelerators that best accommodate each other (Fig.~\ref{fig:co_design_all} (a)).
Software/compiler co-design develops specialized models based on compiler features, and meanwhile customizes compiler instructions for efficient execution (Fig.~\ref{fig:co_design_all} (b)).
Compiler/hardware co-design involves specialized hardware architecture, instruction set extensions, and specialized programming libraries (Fig.~\ref{fig:co_design_all} (c)). 
}

\subsection{Software/Hardware Co-design} 
\label{sec:sw_hw_codesign}

\begin{figure*}
    \centering
    \includegraphics[width=0.9\textwidth]{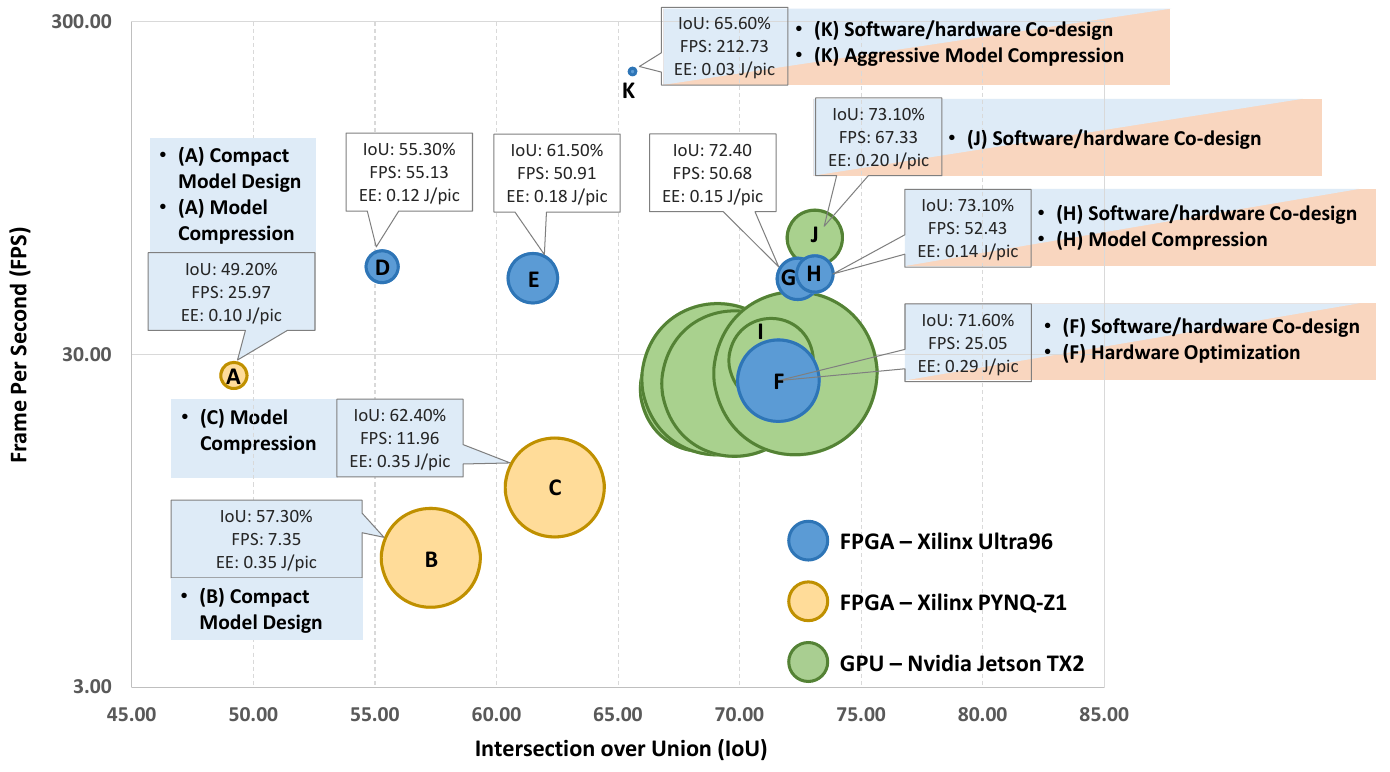}
    \caption{A demonstration on software/hardware co-design in a real-case application from Design Automation Conference System Design Contest (DAC-SDC)~\cite{dacsdc}. The x-axis is the accuracy, intersection-over-Union (IoU); the y-axis is the execution speed, frame-per-second (FPS); the size of the circle represents the energy efficiency, where larger circles mean higher energy per picture.
    (Representative designs: A~\cite{2018fpga2}; B~\cite{2018fpga3}; C~\cite{2018fpga1}; D~\cite{2019fpga3}; E~\cite{2019fpga2}; F~\cite{2019fpga1}; G~\cite{2020fpga3}; H~\cite{2020fpga2}; J~\cite{2019gpu1}; K~\cite{2020fpga1})
    }
    \label{fig:dacsdc_co_design}
\end{figure*}

While hardware-aware NAS approaches (in Sec.~\ref{sec:hardware_aware_NAS}) show promising results, its sequential design methodology limits its effectiveness. During hardware-aware NAS, the hardware accelerator is fixed during NAS but independently optimized afterward. This may result in time-consuming iterations between model and accelerator development, or sub-optimal solutions since the model design does not have sufficient performance feedback from hardware and the hardware deployment does not have a direct impact on model design.
Thus, software models and hardware accelerators show additional benefits from being \textit{simultaneously} designed.

Hao et al.~\cite{hao2018deep} highlight the opportunities of simultaneous software model and hardware accelerator co-design with a prototype solution.
Following this work, Hao et al.~\cite{hao2019fpga} propose the first simultaneous FPGA/DNN co-design framework, including a hardware-oriented bottom-up DNN model design, and a DNN-driven top-down FPGA accelerator design.
A similar work called FNAS, proposed by Jiang et al.~\cite{jiang2019accuracy}, is a hardware-aware NAS framework that can find an optimal neural network under required FPGA implementation latency through reinforcement learning.

NAIS~\cite{hao2019nais} and EDD~\cite{li2020edd} further push co-design to a generalized and unified approach, i.e., fully simultaneous neural architecture and implementation co-search, targeting arbitrary hardware platforms.
NAIS first proposes a stylized design methodology targeting both FPGAs and GPUs, then it takes autonomous driving as a key use case to demonstrate how such a co-design methodology can impact the autonomous driving industry significantly.
Based on NAIS, EDD~\cite{li2020edd} proposes a fully simultaneous, efficient differentiable DNN architecture and implementation co-search methodology. 
Targeting ASICs, Yang et al.~\cite{yang2020co} propose a framework named NASAIC that can simultaneously identify multiple DNN architectures and the associated heterogeneous ASIC accelerators to meet both software accuracy and hardware performance requirements.

\subsubsec{Observations and Insights}
\new{
With the great success of software/hardware co-design, we further provide a quantitative comparison to showcase its effectiveness in solving a real edge AI task:
a light-weight object detection task on drones. The task comes from the Design Automation Conference System Design Contest (DAC-SDC)~\cite{dacsdc}. Contestants propose their own object detection models, implement them on either embedded FPGA or GPU, and evaluate them with respect to a common set of metrics, i.e., accuracy, speed, and energy.}

\new{Fig.~\ref{fig:dacsdc_co_design} summarizes all the solutions including the years of 2018, 2019, and 2020.
The designs located in the bottom-left, such as A, B, and C, are designed by model compression and/or compact model design.
The designs locate in the top-right corner, such as J, K, F, H, employ software/hardware co-design and/or aggressive model compression. 
The figure shows that co-designed models perform better in general. Software/hardware co-design can significantly boost the edge AI solution quality, while applying model compression can further improve the speed and energy efficiency. 
Further, this figure shows that on average, a well-designed FPGA accelerator (e.g., design H) can outperform GPU (e.g., design J) with respect to energy efficiency while often maintaining the same efficiency.}

\subsection{Software/Compiler Co-design} \label{sec:sw-compiler-codesign}

\new{
While software/hardware co-design has been attracting growing research interest, software/compiler co-design remains largely underexplored. This may be partially because designers are more inclined to treat compilers as well-developed tools that should not be touched. Some recent works~\cite{ma2020pconv, niu2020patdnn, liu2020cocopie, ji2020mcunet} address this issue and successfully demonstrate the practicality of software/compiler co-design.}

\new{
Software/compiler co-design includes PCONV~\cite{ma2020pconv}, PatDNN~\cite{niu2020patdnn}, and CoCoPIE~\cite{liu2020cocopie}, which tackle model compression and compilation simultaneously. During model compression, they focus on structured pruning, guided by pre-determined compiler-friendly patterns  (as the example in Fig.~\ref{fig:co_design_all} (b)).
During compilation, they propose efficient compiler code generation, enabling the compilers to maximize or maintain
both instruction-level and thread-level parallelism.
MCUNet~\cite{ji2020mcunet} is another framework that integrates model design and compiler optimization. It is composed of two components, TinyNAS and TinyEngine; TinyNAS searches for specialized DNNs, while TinyEngine generates specialized code to eliminate instruction and memory redundancy.
}

\subsection{Compiler/Hardware Co-design} \label{sec:compiler-hw-codesign}

\new{
Within compiler and hardware co-design, a series of works named PULP~\cite{pulp-main} are among the most important works.
PULP stands for open Parallel Ultra-Low-Power processing platform, and it achieves leading energy-efficiency and widely-tunable performance especially suitable for IoT devices.
As illustrated in Fig.~\ref{fig:co_design_all} (c), the PULP family demonstrates three types of specializations in promoting co-design quality: specialized instruction set architecture (ISA) designs, ISA extensions, and specialized programming libraries.
}

\new{
For specialized processor architecture, there are many chip designs in the PULP family, such as PULP v1 to v3~\cite{rossi201660, rossi2017energy, rossi2017self}, Mr. Wolf~\cite{pullini2018mr}, and GAP-8~\cite{flamand2018gap}.}
\new{For specialized ISA extensions, the work XpulpNN~\cite{garofalo2020xpulpnn} proposes a set of extensions on top of the RISC-V compatible processor architectures for low bit-width quantized DNN executions.}
\new{For specialized programming library, PULP-NN~\cite{garofalo2020pulp} targets a cluster of PULP processors, providing an open-source optimized library to support quantized neural networks.
Similarly, the work FANN-on-MCU~\cite{wang2020fann} provides an open-source toolkit targeting ARM Cortex-M series and PULP platforms.}

\subsubsec{Observations and Insights.}
\new{
Given the great success of compiler/hardware co-design, one can expect that this methodology can be extended to other ASIC architectures and FPGAs. For example, High-Level Synthesis (HLS)~\cite{coussy2009introduction} is a behavior-level compiler and programming tool for FPGA and ASIC, which compiles C/C++ directly to hardware description languages such as Verilog. Currently, both parties, the software developers and the FPGA and ASIC designers, see HLS as a black-box tool. Thus, the final hardware quality largely relies on the quality of HLS tools, and the improvements gained from software and hardware may be canceled due to inefficient code transformation and scheduling inside the HLS compiler. Therefore, it is expected that such compilers should also be co-optimized with specialized hardware architecture in the future.}
\section{conclusions and future directions}  \label{sec:future}

In this paper, we discussed the enabling design methodologies for edge AI spanning the entire development stack, categorized into specialization methodologies and co-design methodologies.
The representative single-layer specialization methodologies we discussed include on-device training algorithms, compact and hardware-aware model design, model compression, adaptive inference, and specialized hardware accelerator designs.
We elaborated three cross-layer co-design methodologies, software/hardware co-design, software/compiler co-design, and compiler/hardware co-design.
Throughout the work, we discussed a few cross-layer co-design opportunities that can boost edge AI solution quality, hoping to inspire future researches in these directions.
Meanwhile, we highlight potential future directions for edge AI, hoping to pave the paths of edge AI development and attract more research interest.

\subsubsec{Robust, Secure, and Private AI.}
As discussed in Sec.~\ref{sec:applications}, the most common design objectives for edge AI are accuracy, latency, energy, and memory, while robustness, security, and privacy issues are still largely underexplored.
AI model training often includes data that requires strict privacy, such as faces, fingerprints, biometrics, financials, and medical information. This introduces risks of attackers injecting malicious data to disrupt edge AI functionality, such as attacks in federated learning~\cite{yang2019federated}.
Likewise, to preserve privacy, the communication between edge and cloud devices needs to be secure. These qualities are necessary for creating practical AI systems that are widely adopted across applications.

\subsubsec{Green, Ethical, and Trusted AI.}

The increasing computational power of AI systems enables higher accuracies yet introduces potential environmental impacts.
Strubel et al.~\cite{strubell2019energy} showed that modern training of AI models results in a significant amount of carbon emissions (e.g., a large NLP model can emit as much carbon as five cars in their lifetimes.).
In addition, existing AI algorithms are vulnerable to biases and errors introduced by human developers or by biased datasets during training.
Datasets sometimes include racial, gender, political, or ideological biases, which may result in wrong, unfair, or discriminatory decisions~\cite{koenecke2020racial}. 
Furthermore, despite the high quality of the existing AI algorithms, the fundamental principles and rationale of the AI outputs are largely unclear to us.
This is problematic in the real world where AI algorithms are making critically important decisions. Developing AI technologies that are environmentally-friendly, fair, trustworthy is essential to the widespread adoption of of successful AI systems.

\textbf{Quantum AI with Q-bits.}
An incredible observation has been made by OpenAI that the amount of compute used in the largest AI training runs has been increasing at an unsustainable rate, doubling every 3.4-months (whereas Moore’s Law had a 2-year doubling period) \cite{ai-compute}.
As a consequence, IBM recently made a statement that future AI systems will require digital bits and Q-bits working in collaboration~\cite{ibm-quantum, quantum-for-ai}.
It has been demonstrated that a superconducting quantum processor is able to perform a traditional machine learning classification task~\cite{quantum-for-ai}, showing significant potential in using quantum computing for machine learning by exploiting the exponentially large quantum state space.
A neural network and quantum circuit co-design framework, QuantumFlow~\cite{jiang2021co}, has pioneered the study of designing neural network models that are suitable for quantum circuits.
Towards this interesting yet challenging new trend of quantum AI, revolutionary design methodologies for software and hardware are required.

\section{Acknowledgments}

This work is supported in part by the IBM-Illinois Center for Cognitive Computing System Research (C3SR) -- a research collaboration as part of IBM AI Horizons Network, by the Semiconductor Research Corporation (SRC) program under Task 2803.001/2804.001, and under NSF Award \#2007832.

\bibliographystyle{IEEEtran}
\bibliography{references}

\end{document}